\documentclass[
aps,
prb,
reprint,
showkeys,
twocolumn,
nofootinbib,
superscriptaddress,
]{revtex4-2}

\usepackage{natbib}
\usepackage{tabularx}
\usepackage{breqn}
\usepackage{textgreek}
\usepackage{subscript}
\usepackage{natbib}
\usepackage{amssymb}
\usepackage{color}
\usepackage{amsfonts}
\usepackage{textcomp}
\usepackage{amsmath}
\usepackage[charter]{mathdesign}
\usepackage{easyReview}
\usepackage{microtype} 
\usepackage{upgreek}
\usepackage{subfigure}
\usepackage{booktabs}
\newcommand{\ped}[1]{\ensuremath{_{\rm #1}}}

\definecolor{link}{RGB}{57,106,177}
\definecolor{darkgreen}{RGB}{0,128,0}
\definecolor{blue}{RGB}{0,0,0}
\usepackage{float}
\usepackage{graphicx}
\usepackage{hyperref}

\hypersetup{
	colorlinks=true,
	linkcolor=link,
	citecolor=link,
	urlcolor=link
}

\begin{document}

\title{Sub-shot-noise emission statistics of a CW-excited single photon source}
	
\author{G. Gavello}
\affiliation{Physics Department - University of Torino - Torino 10025 - Italy}

\author{G. Petrini}
\affiliation{Physics Department - University of Torino - Torino 10025 - Italy}
\affiliation{Istituto Nazionale di Ricerca Metrologica
(INRiM) - Torino 10135 - Italy}

\author{I. Ruo Berchera}
\affiliation{Istituto Nazionale di Ricerca Metrologica
(INRiM) - Torino 10135 - Italy}

\author{M. Gramegna}
\affiliation{Istituto Nazionale di Ricerca Metrologica
(INRiM) - Torino 10135 - Italy}

\author{E. Moreva}
\affiliation{Istituto Nazionale di Ricerca Metrologica
(INRiM) - Torino 10135 - Italy}

\author{P. Traina}
\affiliation{Istituto Nazionale di Ricerca Metrologica
(INRiM) - Torino 10135 - Italy}
\author{M. Ziino}
\affiliation{Physics Department - University of Torino - Torino 10025 - Italy}
\author{M. Genovese}
\affiliation{Istituto Nazionale di Ricerca Metrologica
(INRiM) - Torino 10135 - Italy}
\affiliation{Istituto
Nazionale di Fisica Nucleare (INFN) - Sezione di Torino -
Torino 10125 - Italy}
\author{P. Olivero}
\affiliation{Physics Department - University of Torino - Torino 10025 - Italy}
\affiliation{Istituto
Nazionale di Fisica Nucleare (INFN) - Sezione di Torino -
Torino 10125 - Italy}
\author{J. Forneris}
\affiliation{Physics Department - University of Torino - Torino 10025 - Italy}
\affiliation{Istituto
Nazionale di Fisica Nucleare (INFN) - Sezione di Torino -
Torino 10125 - Italy}
\affiliation{contact: jacopo.forneris@unito.it}
\author{I.P. Degiovanni}
\affiliation{Istituto Nazionale di Ricerca Metrologica
(INRiM) - Torino 10135 - Italy}
\affiliation{Istituto
Nazionale di Fisica Nucleare (INFN) - Sezione di Torino -
Torino 10125 - Italy}
\affiliation{contact: i.degiovanni@inrim.it}


\begin{abstract}
\noindent Shot noise sets a fundamental limit on the sensitivity of classical optical measurements, with coherent emitters achieving the lowest possible shot-noise level. Emission from sub-Poissonian light provides a pathway to surpass this limit, and single-photon sources provide a natural platform for generating such light.  However, it is commonly assumed that off-resonance continuously excited solid-state  single-photon sources exhibit Poissonian statistics over large exposure periods, owing to their strong coupling to the host lattice. In this work, a theoretical model of a continuously driven two-level single-photon source is developed, treating both excitation and radiative decay as stochastic processes. The analysis demonstrates that photon emission can display sub-Poissonian statistics when excitation and decay rates are comparable, showing that continuous excitation does not inherently preclude nonclassical emission. The model is further extended to include finite detection efficiency and detector dead time, illustrating how these practical non-idealities can affect the experimental observation of sub-Poissonian statistics.
\end{abstract}

\maketitle
\noindent
\section{Introduction}
\noindent Optical measurements are fundamentally limited by shot noise, a consequence of the quantized nature of light. \\
In classical imaging, the lowest shot-noise level is achieved using coherent light, which follows Poissonian photon statistics. Light sources with larger intensity fluctuations -- such as thermal light, which exhibits super-Poissonian statistics -- produce even higher level of noise. Consequently, the Poissonian case sets a strict upper bound on the signal-to-noise ratio achievable in classical imaging. Enhancing measurement precision within this limit requires either increasing optical power or extending acquisition times, approaches that may be impractical in low-light conditions or potentially damaging to sensitive samples.

\noindent Quantum imaging, by contrast, offers a way to surpass these classical constraints by exploiting light with nonclassical photon statistics \cite{gattomonticone2014beating, schwartz2012improved, tenne2019super, berchera2019quantum, genovese2016real,  picariello2025quantum, Lawrie2019QuantumSensing, Defienne2024AdvancesQuantumImaging, Pirandola2018AdvancesPhotonicSensing, Dorfman2016NonlinearOpticalSignals}. In particular, sub-Poissonian light, where photon-number fluctuations are suppressed below the Poissonian level, can improve measurement precision without increasing the mean photon number. A common implementation involves using twin-beam states \cite{brambilla2008high, brida2010experimental, brida2011experimental, ortolano2023quantum}, generated through spontaneous parametric down-conversion or four-wave mixing. In these states, a single pump photon produces two strongly correlated photons: the signal photon, which probes the object, and the idler photon, which serves as a reference. By exploiting their quantum intensity correlations and performing differential detection, correlated fluctuations can be canceled, allowing precision beyond the classical shot-noise limit \cite{samantaray2017realization, Ortolano2023pattern, sabines2019twin, Wang2024advantage}.While progresses have been made towards applicability with twin beam strategies \cite{ruo2020improving, paniate2026high}, to achieve high spatial resolution imaging is still a challenge due to the intrinsic coherence size of  quantum correlations and the need of preserving through high numerical aperture optical system.
\\
An alternative approach involves using single-photon sources (SPSs), which provide a more direct path to sub-Poissonian statistics. Ideally, a SPS emits exactly one photon at a time on demand, with each emitted photon being identical and indistinguishable from the others \cite{lounis2005single, EsmannWeinAntonSolanas2024SolidState, Mohammadnejad2023RoomTemperature, Holmes2020QuantumDotSources, Senellart2017HighPerformanceQD, Andrini2024SolidStateColorCenters, Castelletto2021SiCSources, TothAharonovich2019AtomicallyThin, Zhang2024TwoDimensionalEmitters, Ye2024ColloidalQDs, Rempe2007CavityQED, DuanMonroe2010TrappedIonNetworks,  Toninelli2021SingleMolecules, ShkarinGotzinger2026Molecules}. Various platforms have been developed to realize such emitters. 
Atomic and ionic systems, including trapped atoms or ions \cite{higginbottom2016pure}, deliver extremely pure single-photon emission with well-defined quantum states, though they typically require complex trapping and cooling.
Solid-state systems, such as diamond color centers \cite{gattomonticone2014native, corte2022magnesium} and semiconductor quantum dots \cite{costa2026quantum}, combine high brightness with on-chip integration \cite{palm2023modular}, with diamond centers operable  off-resonance at room temperature. 
Emerging platforms, such as defect-engineered non-diamond crystals \cite{grosso2017tunable, radulaski2017scalable} and two-dimensional materials such as transition metal dichalcogenides \cite{parto2021defect}, further expand the options for scalable and versatile SPS implementation. \\
Despite their potential as sub-shot noise emitters, it is generally well known that off-resonance continuous-wave pumped solid-state SPSs exhibit the typical exponential decay  of the spontaneous emission that leads to Poissonian statistics over long integration periods (i.e. much longer than the the typical excited state lifetime of the spontaneous emission), due to their strong coupling with the host lattice rendering them unsuitable for sub-shot-noise applications that require sub-Poissonian behavior \cite{berchera2019quantum, genovese2016real, Lawrie2019QuantumSensing, Defienne2024AdvancesQuantumImaging, Pirandola2018AdvancesPhotonicSensing, Dorfman2016NonlinearOpticalSignals}. In this work, theoretical modeling of a continuously driven two-level SPS demonstrates that, under specific conditions, such systems can exhibit sub-Poissonian photon statistics. This indicates that continuous excitation does not inherently preclude nonclassical emission. Based on this finding, the feasibility of experimental validation is examined, accounting for the influence of realistic non-idealities on the detection of sub-Poissonian behavior.
\\
\section{Model}
The system addressed in this model is a solid state emitter in which photon emission occurs only via de-excitation from an excited state S2 to a ground state S1. The availability of non-radiative de-excitation paths or system inter-crossing via shelving states \cite{Berthel2015Photophysics, Santori2004SinglePhoton, Demissie2020TripletShelving} is neglected in the following discussion. 
\\
The system is initially prepared in its ground state and subsequently exposed to continuous off-resonance optical excitation. Upon absorption of a pump photon, the SPS is promoted to higher vibrational states of S2, where ultra-fast internal conversion processes, related to the interaction with the surrounding crystal lattice \cite{Fox2010OpticalProperties, Alkauskas2014Luminescence}, convey the system to S2. The system then relaxes back to the ground level following the usual exponential decay process of the spontaneous photon emission. The internal conversion mechanism has the practical effect of introducing rapid decoherence in the Rabi oscillations induced between S1 and S2 by the excitation field. 
Thus, the probability density function (PDF) describing the spontaneous emission of a single-photon, assuming that the SPS reaches the excited state at t=0 is, as said above  
\begin{equation}
f\ped{em}(t) = \mu_2 e^{-\mu_2\,t}\theta(t).
\label{eq:fem}
\end{equation}
where $1/\mu_2$  is the typical excited state lifetime of the spontaneous emission ($\tau$), and  $\theta$ is the Heaviside function.
Therefore, the resonant model first introduced by L. Mandel \cite{Mandel1979SubPoissonian, Hodapp1990} is not relevant in this context. Also, the excitation mechanism can be regarded as a stochastic process described by an exponential behavior typical of the Poissonian process, i.e. assuming that the optical pump is activated at t=0, the probability density function (PDF) describing the absorption of a photon is:     
\begin{equation}
f\ped{abs}(t) = \mu_1 e^{-\mu_1\,t}\theta(t),
\label{eq:fabs}
\end{equation}
where $1/\mu_1$ describes the typical time, dependent on the exciting pump power, required by the system to to reach the S2 level.

Within this framework, assuming that each subsequent absorption–emission cycle is independent of the others, the PDF for the SPS to emit a photon given the optical pump activated at $t=0$ can be expressed as the following convolution of the two PDFs presented in eqs.\ref{eq:fabs} and \ref{eq:fem}:
\begin{equation}
\nonumber
f(t) = \int dt'dt''f\ped{abs}(t')f\ped{em}(t'')\,\delta(t-t'-t''),
\end{equation}
whose analytical expression is: 
\begin{equation}
f(t) = \frac{\mu_1\mu_2}{\mu_1-\mu_2} \left( e^{-\mu_2\,t} - e^{-\mu_1\,t} \right) \theta(t).
\label{eq:fSPS}
\end{equation}
From the expression above, it is evident that in the limiting cases where $\mu_1 \gg \mu_2$ or $\mu_1 \ll \mu_2$, the statistics of the combined absorption–emission process is Poissonian, reflecting the fact that the dominant process itself follows Poisson statistics. In contrast, in the intermediate regime the behavior is not immediately apparent and the moments of the corresponding statistical distribution must therefore be explicitly derived. In principle, this would involve determining the probability $P_T(n)$ that the SPS emits $n$ photons within a time interval $T$ and subsequently calculating the moments from $P_T(n)$. However, this approach is not practical here, as it would require evaluating a $n$-fold convolution of PDFs of the form given in Eq.\,(\ref{eq:fSPS}), corresponding to an ordered sequence of excitation and subsequent spontaneous decay of a single emitter (see appendix A for details). A more convenient approach is to introduce the probability generating function (PGF). \\
For $n$, the number of photons emitted by the SPS during the time interval $T$, the PGF is defined as $G(T,\xi)= \sum_{n=0}^\infty P_T(n) \xi^n $. This function is a powerful tool, as it enables the moments of the photon statistics to be straightforwardly derived from its analytical expression via simple differentiation with respect to $\xi$. Although the convolution issue formally persists -- since $P_T(n)$ corresponds to a $n$-fold convolution of PDFs as discussed in appendix A -- it can be overcome by considering the Laplace transform of the PGF: $\tilde{G} (s, \xi)$. As shown in Appendix B, this quantity takes the form 
\begin{equation}
\tilde{G} (s, \xi)= \frac{1}{s} \frac{1- \tilde{f}(s)}{1-\xi \tilde{f}(s)},
\label{eq:Gtilde}
\end{equation}
where $\tilde f (s)$ is the Laplace transform of a single PDF of the type given in eq.\,\ref{eq:fSPS}. Explicitly:
\begin{equation}
\tilde{f} (s)= \frac{\mu_1 \mu_2}{(\mu_1+s)\,(\mu_2+s)}.
\end{equation}
\noindent From eq.\,\ref{eq:Gtilde}, the moments of the photon statistics are obtained by differentiating $\tilde{G}(s,\xi)$ with respect to $\xi$ and then applying the inverse Laplace transform to return to the time domain. For instance, the mean number of photons emitted by the SPS over $T$, denoted $\langle n \rangle_T$, is obtained as:
\begin{equation} \nonumber
\langle n \rangle_T = \mathscr{L}^{- 1} \left[ \frac{d}{d\xi}\tilde{G} (s, \xi) \bigg|_{\xi=1} \right],
\end{equation}
where $\mathscr{L}^{-1}$ represents the inverse Laplace transform operation. Similarly, the variance $Var[n]_T$ is obtained as $Var[n]_T= \langle n(n-1)\rangle_T +\langle n\rangle_T -\langle n\rangle^2_T $, with $\langle n(n-1) \rangle_T$ calculated via
\begin{equation} \nonumber
\langle n(n-1) \rangle_T = \mathscr{L}^{- 1} \left[ \frac{d^2}{d\xi^2}\tilde{G} (s, \xi) \bigg|_{\xi=1} \right].
\end{equation}

\begin{figure}[h]
  \includegraphics[width= 0.95\linewidth]{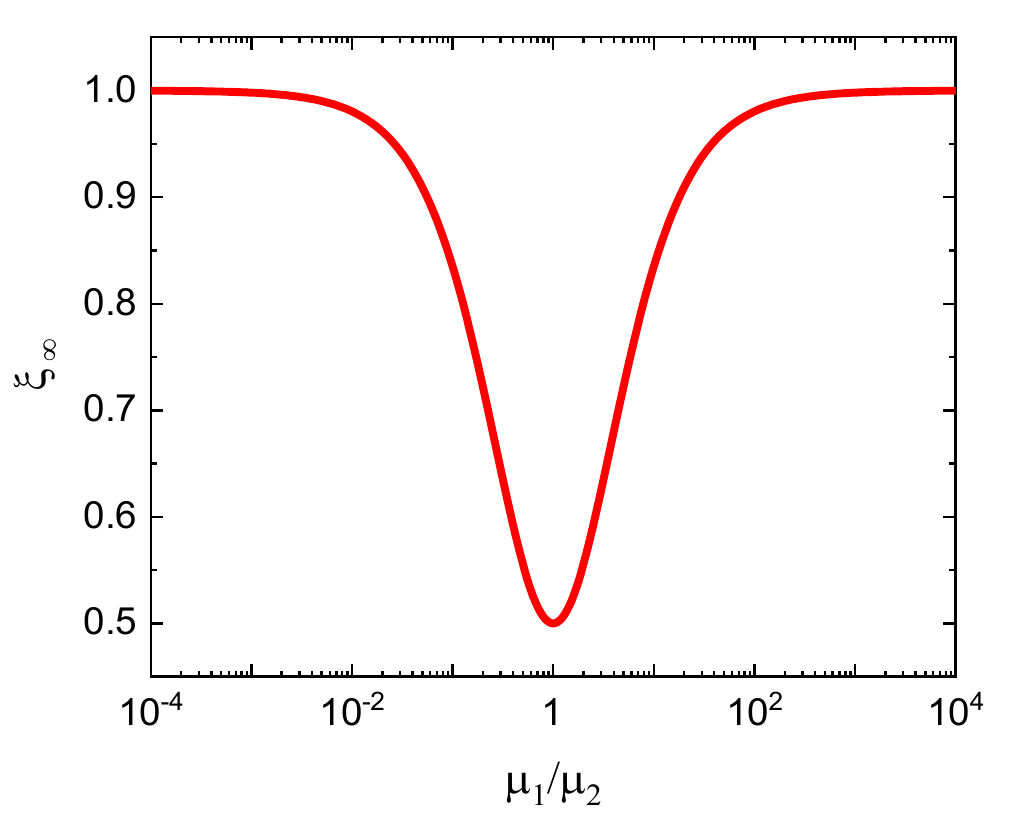}
  \caption{Plot of the quantity $\xi_\infty$, defined as the ratio of the variance to the mean number of emitted photons by a continuously excited single-photon source in the large-time limit, as a function of the ratio $\mu_1$/$\mu_2$.}
  \label{fig:VarN/N_noloss}
\end{figure}

\noindent The quantities $\langle n \rangle_T$ and $Var[n]_T$ are obtained following the procedure described above, with their general expressions reported in Appendix C. Only their asymptotic forms for long times ($T \to \infty$) are considered here, as photon counting is usually carried out over macroscopic time windows ranging from ms to s, which can be treated as effectively infinite compared to the ns/$\mu$s relaxation times typical of existing SPSs \cite{sipahigil2014indistinguishable, storteboom2015lifetime}. In this limit, the quantities are given by:  \begin{equation}
\langle n \rangle_{\infty}  = \frac{\mu_1\mu_2}{(\mu_1+\mu_2 )}  T
\label{eq:Ninfid}
\end{equation}
\begin{equation}
Var[n]_{\infty}=  \frac{\mu_1^2 + \mu_2^2}{(\mu_1+\mu_2 )^2} \cdot \langle n \rangle_{\infty}
\label{eq:VarNinfid}
\end{equation}
At this stage, it is convenient to introduce the quantity $\xi_\infty = Var[n]_\infty/ \langle n \rangle _\infty$, known as the Fano factor \cite{Fano1947} which provides a direct measure of the emission statistics. The Fano factor is simply related to the well known Mandel's quantumness quantifier $Q$ \cite{Mandel1979SubPoissonian} as $Q=\xi -1$.  Figure \ref{fig:VarN/N_noloss} shows the Fano factor  $\xi_\infty$ as a function of $\mu_1/\mu_2$. 

\noindent In agreement with the calculations above, the statistics is poissonian ($\xi_\infty = 1$) when $\mu_1 \gg \mu_2$ or $\mu_1 \ll \mu_2$. Interestingly, in the intermediate regime, where no prior information was available, the emission statistics turns out to be sub-poissonian ($\xi_\infty <1$), reaching $\xi_\infty = 1/2$ for $\mu_1=\mu_2$. \\

\noindent To complete the characterization of the SPS, its asymptotic emission intensity is directly obtained from eq.\ref{eq:Ninfid} as the mean number of photons emitted over the macroscopic measurement time $T$. Using the relationships linking $\mu_1$ and $\mu_2$ to the pump power $P$ and the characteristic lifetime $\tau$ of the source, the photon rate is expressed as:
\begin{equation*}
I_\infty =  \frac{1}{\tau} \frac{\alpha P}{\alpha P+ 1/\tau}      
\end{equation*}
which is commonly rewritten in the familiar saturation form:
\begin{equation*}
I_\infty = I_{sat} \frac{P}{P+P_{sat}}   
\end{equation*}
where $I_{sat} = \tau^{-1}$ is the maximum photon emission rate of the SPS, and the saturation power is $P_{sat} = (\tau\, \alpha)^{-1}$.

\section{Discussion}
\noindent 
\noindent The identification of a regime in which a continuously excited SPS exhibits sub-poissonian emission is particularly significant, as it points to the possibility of achieving sub-shot-noise imaging with these systems. Any realistic evaluation of this potential, however, requires experimental verification of the regime predicted. This would involve measuring $\xi_\infty$ at different laser powers to span different values of $\mu_1$, while $\mu_2$ remains fixed by the choice of SPS. Observing a minimum in $\xi_\infty$ when $\mu_1 = \mu_2$ would thus indicate the sub-Poissonian nature of the continuously excited SPS. \\
However, it is essential to remark that practical photon-counting measurements are affected by inefficiencies at the collection level. As a result, the statistics of the detected photons may differ from the intrinsic emission statistics of the SPS, causing the observed behavior to deviate from the idealized model described above. To address this, the model is extended to analyze how non-idealities influence the outcome and to identify the conditions under which a sub-Poissonian signature can still be observed.

\noindent Initially, the focus is on optical losses that can be quantified by a detection efficiency $\eta$, defined as the ratio of photons detected to photons emitted by the SPS. This parameter can encompass a variety of loss mechanisms, including imperfect photon extraction from the SPS, suboptimal coupling between the source and the detection apparatus, the finite efficiency of the detector itself. Furthermore, $\eta$ can be extended to account for the fact that in real SPSs, not every photon absorbed leads to a radiative emission \cite{nguyen2019photodynamics}, as alternative relaxation pathways may exist. \\
Within this framework, if the SPS emits $n$ photons during a measurement interval $T$, only a fraction $\eta$ is registered by the measurement system. Consequently, the detection PDF is not simply the convolution of $n$ copies of the SPS $f(t)$ from eq.\,\ref{eq:fSPS}, denoted by $f\ped{n}(t)$ (see Appendix A). To illustrate this point, consider the scenario in which the $n$-th emitted photon is the first one to be detected, while the preceding $n-1$ photons were missed due to imperfect detection efficiency. In this case, the corresponding PDF is given by $\eta\,(1-\eta)^{n-1} f_n(t)$.

\begin{figure}[h]
  \includegraphics[width= 0.95\linewidth]{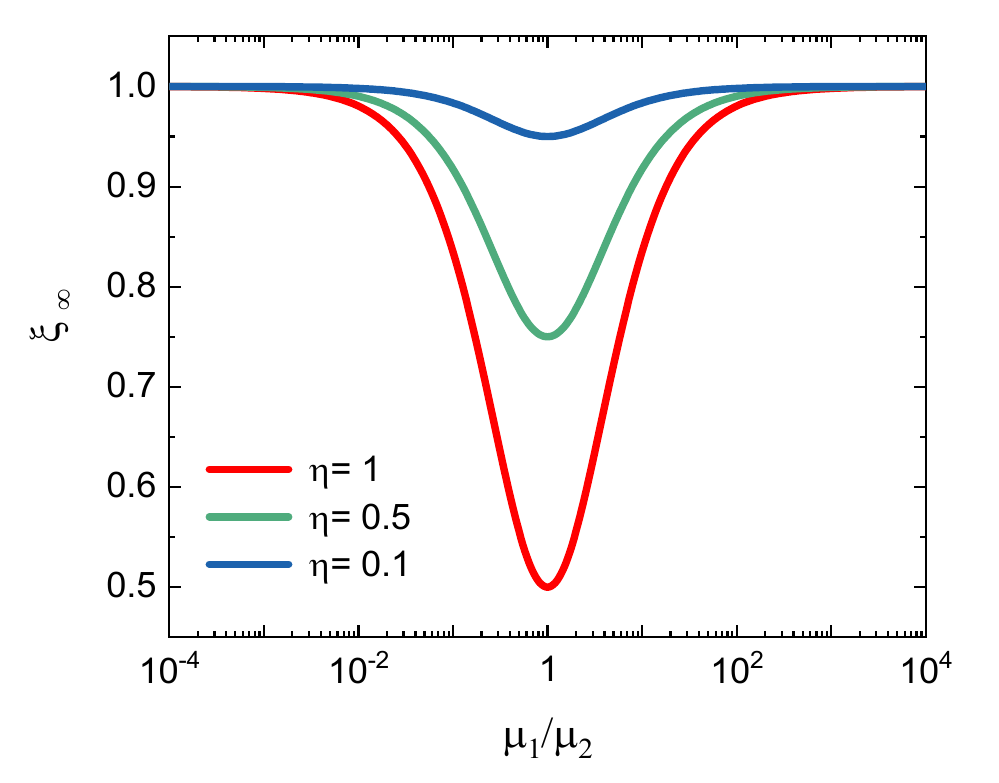}
  \caption{Plot of the quantity $\xi_{\infty} (\eta)$, defined as the ratio of the variance to the mean number of emitted photons by a continuously excited single-photon source in the large-time limit in the presence of optical losses. Losses are quantified by the detection efficiency $\eta$. The ratio is displayed as a function of $\mu_1/\mu_2$, with each colored curve corresponding to a different $\eta$ value indicated in the legend.}
  \label{fig:VarN/N_loss}
\end{figure}

\noindent Generalizing this idea, the total detection PDF must account for all possible scenarios in which an arbitrary number of photons are lost before a successful detection occurs. This leads to:
\begin{equation}
f_\eta (t)= \sum_{n=1}^\infty \eta\,(1-\eta)^{n-1}f_n(t),
\label{eq:fetat}
\end{equation}
whose Laplace transform, as shown in Appendix D, is given by: 
\begin{equation}
\tilde{f}_\eta (s)= \eta\,\frac{\tilde{f}(s)}{1-(1-\eta)\tilde{f}(s)}.
\label{eq:fetas}
\end{equation}
\noindent Once this expression is obtained, the statistics at the detection stage is evaluated by following the same procedure previously adopted. Specifically, the Laplace transform of the PGF is computed by replacing $\tilde{f}(s)$ in eq.\,\ref{eq:Gtilde} with the modified function $\tilde{f}_{\eta}(s)$, thereby explicitly accounting for optical losses. From the resulting expression, the mean number of detected photons and their variance in the presence of losses, $\langle n_\eta \rangle_T$ and $Var[n_\eta]_T$, are determined through differentiation with respect to the generating variable, followed by inverse Laplace transformation. Restricting the attention to the large-time limit ($T\rightarrow \infty$) as done before, their asymptotic expressions are given by:
\begin{equation} \nonumber
\langle n_\eta \rangle_{\infty}  = \eta \frac{\mu_1\mu_2}{(\mu_1+\mu_2 )}  T 
\label{eq:Ninfloss}
\end{equation}
\begin{equation} \nonumber
Var[n_\eta]_{\infty}=  \frac{\mu_1^2 + 2(1-\eta)\mu_1\mu_2 + \mu_2^2}{(\mu_1+\mu_2 )^2} \cdot \langle n_\eta \rangle_{\infty}
\label{eq:VarNinfloss}
\end{equation}
In this case as well, it is convenient to compute the ratio between these former quantities, namely $\xi_{\infty} (\eta) = Var[n_\eta]_\infty/ \langle n_\eta \rangle _\infty$. Figure \ref{fig:VarN/N_loss} displays this quantity as a function of $\mu_1/\mu_2$ for three different detection efficiencies: $\eta = 1$ (ideal reference case, red solid line), $\eta = 0.5$ (green solid line), and $\eta = 0.1$ (blue solid line). When compared with the ideal scenario, it becomes clear that detection losses produce a uniform vertical rescaling of the curve. Although the shape and the position of the minimum remain unaffected, its depth is progressively reduced as the efficiency decreases. Thus, the nonclassical signature encoded in the suppression of $\xi_{\infty}$ at $\mu_1=\mu_2$ becomes less pronounced for smaller $\eta$.

\noindent 
Experimentally, this means that tuning $\mu_1$ by sweeping the pump power (with $\mu_2$ fixed by the selected SPS) will yield a flatter profile than in the lossless case. For very low efficiencies, the contrast of the minimum may approach the level of instrumental noise, increasing the risk that the sub-Poissonian nature of the SPS becomes partially or completely obscured in realistic measurements.\\
\noindent Attention is now directed to another non-ideality affecting the detection process that must be accounted for in a realistic discussion, namely the finite dead time of detectors. Depending on the underlying physical mechanisms, detectors are generally classified as either paralyzable or non-paralyzable. In the non-paralyzable case, once an event is registered, the system becomes inactive for a fixed interval -- referred to as the dead time and denoted by $D$. During this period, any subsequent incoming events are simply ignored and do not extend the duration of the inactive state. By contrast, in paralyzable detectors, events occurring within the dead time prolong its duration. Restricting the discussion to non-paralyzable systems -- such as single-photon avalanche diodes (SPADs) operated in Geiger mode with active quenching circuits, which constitute the most widely used commercial solutions for single-photon detection \cite{georgieva2021detection} -- the average detected
photons in the measurement time $T$ ($\langle n_{out} \rangle$) is related to the incoming Poissonan rate $\nu_{in}$ by:
\begin{equation*}
\langle n_{out} \rangle = \frac{\nu_{in}}{1+D\,\nu_{in}},
\end{equation*}
and the associated uncertainty is $Var[n_{out}]=\langle n_{out} \rangle /(1+D\,\nu_{in})^2 $.
\\
These equations indicate that the measured rate ($<n_{out}>/T$) is reduced with respect to the true rate $\nu_{in}$ and that the associated uncertainty shows sub-Poissonian behavior; these behaviors depend on both the dead time $D$ the incoming photon rate. We note that these results are obtained by modifying the PDF of the Poissonian process accounting for the presence of the dead time $D$, i.e. $f(t)= A \exp(-\nu_{in} t ) \theta (t-D)$ \cite{cox1962renewal}.  

\begin{figure}[h]
  \includegraphics[width= 0.95\linewidth]{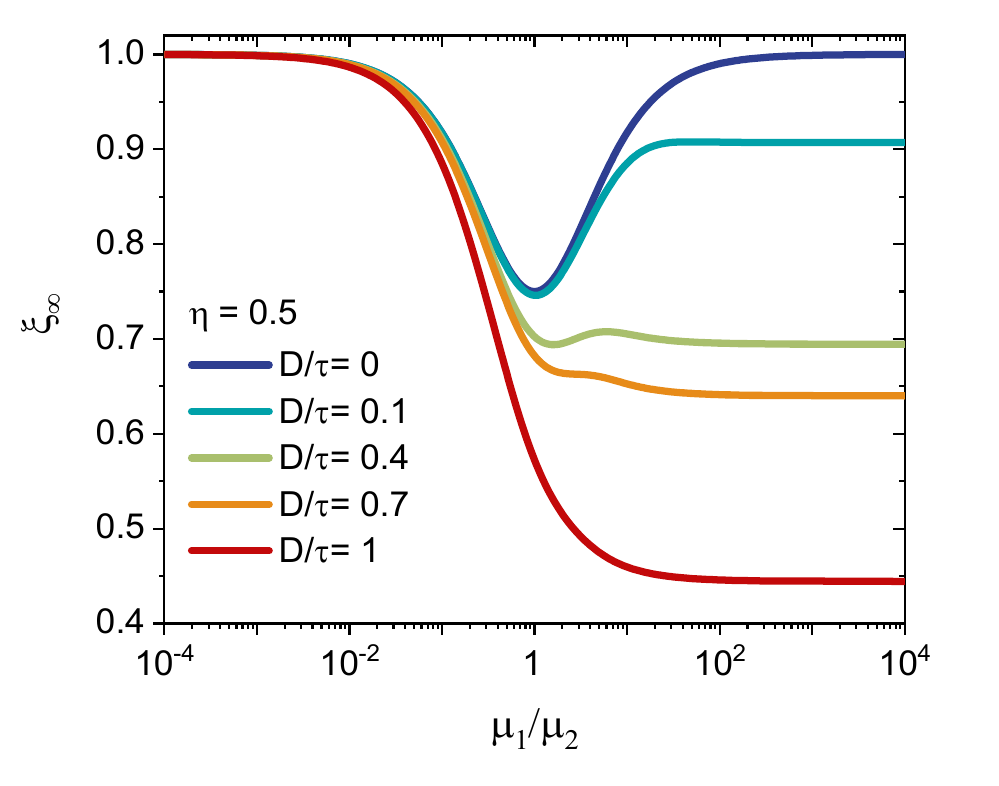}
  \caption{Plot of the quantity $\xi_{\infty} (D, \eta)$, defined as the ratio of the variance to the mean number of emitted photons by a continuously excited single-photon source in the large-time limit in the presence of optical losses and finite detector dead time $D$. The ratio is displayed as a function of $\mu_1/\mu_2$, with each colored curve corresponding to a different $D$ value indicated in the legend, expressed in unit of the characteristic lifetime $\tau$ of the single-photon source. Since the impact of optical losses has already been considered, all curves are calculated for a fixed optical efficiency: $\eta=0.5$.}
  \label{fig:VarN/N_DD}
\end{figure}

Consequently, the detection PDF derived in eq.\,\ref{eq:fetat} no longer accurately represents the statistics in the presence of a non-zero detector dead time. In this case, the correct PDF must instead be written as $f\ped{D,\eta} (t)= A\,f_\eta(t) \,\theta(t-D)$ so that, given a detection event at $t=0$, the distribution remains zero throughout the detector recovery interval and, after the dead time $D$ has elapsed, becomes proportional to $f\ped{\eta}(t)$. The constant $A$ ensures proper normalization of the modified PDF. Enforcing this normalization condition yields the expression:
\begin{equation} \nonumber
f\ped{D,\eta} (t)= \frac{e^{-\frac{\mu_{m}t}{2}} - e^{-\frac{\mu_{p}t}{2}}}{2 \left( \frac{e^{-\frac{\mu_{m}D}{2}}}{\mu_m} - \frac{e^{-\frac{\mu_{p}D}{2}}}{\mu_p}\right)} \theta(t-D) 
\end{equation} 
with $\mu_{p}= (\mu_1+\mu_2) + \sqrt{(\mu_1+\mu_2)^2-4 \eta \mu_1 \mu_2} $ and $\mu_{m}= (\mu_1+\mu_2) - \sqrt{(\mu_1+\mu_2)^2-4 \eta \mu_1 \mu_2} $. Following the same procedure as before, the Laplace transform of the PDF just obtained is computed, resulting in:
\begin{equation} \nonumber
\tilde{f}_{D, \eta}(s)= \frac{\mu_p \mu_m \left( \frac{2 e^{- \frac{1}{2} (2s+\mu_m)D}}{2s+\mu_m} - \frac{2 e^{- \frac{1}{2} (2s+\mu_p)D}}{2s+\mu_p}\right)}{2 \left( \mu_p e^{-\frac{\mu_mD}{2}}- \mu_m e^{-\frac{\mu_p D}{2}} \right)}.
\end{equation}
The Laplace transform of the PGF is then computed by replacing $\tilde{f}(s)$ in eq.\,\ref{eq:Gtilde} with the modified function $\tilde{f}_{D, \eta}(s)$, thereby explicitly accounting for both optical losses and dead time. From this expression, the next step would involve determining the mean number of detected photons and their variance, $\langle n_{D,\eta} \rangle_T$ and $Var[n_{D,\eta}]_T$, through differentiation with respect to the generating variable, followed by inverse Laplace transformation. In this case, however, the inverse Laplace transform does not yield a closed-form analytical expression as it did for zero dead time. This difficulty can nevertheless be circumvented, since the analysis only requires the asymptotic limit for long acquisition times ($T\rightarrow \infty$), for which an analytical function can still be derived. Concretely, this is achieved by retaining only the terms in the series expansion of the Laplace-transformed PGF whose order is negative and then performing the inverse transformation. This procedure is justified by the fact that the limit $T\rightarrow \infty$ in the time domain corresponds to $s \rightarrow 0 $ in the Laplace one. Due to the length and complexity of the resulting expressions, the explicit forms of $\langle n_{D,\eta} \rangle_\infty$ and $\mathrm{Var}[n_{D,\eta}]_\infty$ are not reported here. Instead, their ratio in the asymptotic limit, defined as $\xi_{\infty} (D, \eta)= Var[n_{D,\eta}]_\infty/ \langle n_{D,\eta} \rangle _\infty$ , is presented in figure \ref{fig:VarN/N_DD}. Since the impact of optical losses has already been examined,  $\xi_{\infty} (D,\eta)$ is evaluated at a fixed optical efficiency $\eta=0.5$. The dead time is expressed in units of the characteristic lifetime $\tau$ of the SPS and is varied from $D=0$\,s (dark blue solid line) to $D=\tau$ (red solid line).\\

\noindent Unlike the previous case, a finite dead time has a more surprising impact on $\xi_\infty$, progressively distorting the curve as the dead time $D$ increases. Even at a relatively short dead time of $D = 0.1 \tau$ (solid light blue line), a clear asymmetry emerges: the plateau at $\mu_1 \gg \mu_2$ drops below 1, whereas the plateau at $\mu_2 \gg \mu_1$ remains largely unaffected. This asymmetry can be consistently interpreted as an artificial sub-Poissonian signature induced by the non-paralyzable dead time. Indeed, in the regime $\mu_1 \ll \mu_2$, the dynamics is dominated by a slow absorption time, which effectively limits the counting rate. As a result, detection events are sufficiently sparse that the dead time plays a negligible role, and the system reproduces the behavior expected in the zero-dead-time limit. Conversely, when $\mu_1 \gg \mu_2$, the intrinsic event rate is significantly higher, and the dead time actively suppresses closely spaced detection events,  essentially forcing an antibunching behavior. This leads to an apparent sub-poissonian character and, consequently, to a lowering of the $\xi_\infty$ plateau. As the dead time increases further, the asymmetry becomes more pronounced: at  around $D = 0.4 \tau$ (green solid line), the position of the minimum shifts slightly away from $\mu_1 = \mu_2$. When the dead time approaches the SPS lifetime ($D = \tau$, red solid line), the minimum disappears entirely, and the measurement is completely dominated by the fake sub-poissonian  behavior induced by the detector dead time, where the detected events are typically temporally separated by $D$ at least. This is the well known saturation behavior single-photon detectors with non-paralyzable dead time whose maximum count rate is (ideally) $1/D$ \cite{cox1962renewal}. From an experimental standpoint, this highlights the importance of using detectors with a dead time shorter than the lifetime of the SPS under investigation. Otherwise, the sub-poissonian character of the source may be confused with non-idealities of the detection system.

\section{Conclusions}
\noindent
In this work, a comprehensive theoretical model was developed for a continuously excited two-level single-photon source, treating excitation and radiative decay as stochastic processes within a unified description. By employing the probability generating function and its Laplace-transform properties, explicit expressions for the long-time mean photon number and variance were obtained, providing a direct characterization of the emission statistics.\\
The analysis shows that the single-photon source exhibits Poissonian statistics when one of the underlying dynamical processes dominates. When excitation and radiative decay occur on comparable time scales, the photon statistics become sub-Poissonian, demonstrating that continuous excitation does not preclude intrinsically nonclassical emission.\\
\noindent The framework was further extended to include realistic detection effects. Limited collection efficiency reduces the visibility of the nonclassical signature without fundamentally altering its structure, whereas finite detector dead time can significantly distort the statistics and may induce a fake sub-Poissonian character.\\
\noindent Overall, the theoretical framework developed here not only elucidates the statistical properties of continuously excited single-photon sources but also provides practical guidance for experimental verification and the development of quantum-enhanced technologies.

In this perspective it is important to note that the sub-Poissonian behavior presented here is strongly degraded by the presence of non-unit detection efficiency and optical losses, as often happens in technological applications exploiting sub-Poissonian light \cite{berchera2019quantum, genovese2016real, Pirandola2018AdvancesPhotonicSensing}. In order to reliably measure the phenomenon it is important to minimize the optical losses and a single-photon detector with high detection efficiency and a dead time which is much shorter than the spontaneous decay time of the emitter is required.

\section*{Appendix A}
\noindent As outlined in the main text, $P_T(n)$ involves a $n$-fold convolution of PDFs of the form given in eq.\,\ref{eq:fSPS}, which renders the direct calculation of the corresponding statistical moments impractical. The origin of this $n$-fold convolution is clarified here. Specifically, $P_T(n)$ can be expressed in terms of the auxiliary quantity $K\ped{n}(T)$, defined as the probability that the $n$-th photon emitted by the SPS occurs within the time interval $[0, T]$. From this definition, this quantity is given by the integral over $[0,T]$ of the convolution of $n$ copies of eq.\,\ref{eq:fSPS}:
\begin{equation*}
K_n (T)=\int_0^T f(t_1 )…f(t_n )  \delta(t-t_1…-t_n) dt dt_1…dt_n	
\label{eq:Kn}
\end{equation*}
which can be rewritten more compactly as
\begin{equation}
K_n(T) = \int_0^T f_n(t) dt
\label{eq:Kn}
\end{equation}
where $t= \sum_{i=1}^n t_i$ arises from the action of the Dirac delta function, and $f_n(t)$ denotes the $n$-fold convolution of $f(t)$.\\

\noindent Equivalently, $K_n (T)$ can be interpreted as the cumulative probability of having at least $n$ photons emitted within the same interval. This leads to the relation
\begin{equation*}
K_n (T)=\sum_{n'=n}^\infty P_T (n' )
\end{equation*} 
from which, by straightforward manipulation, one obtains: 
\begin{equation}
P_T (n)=K_n (T)-K_{n+1} (T)
\label{eq:PnKn}
\end{equation}
This relation explains the appearance of the $n$-fold convolution of PDFs in the expression for $P_T(n)$.

\section*{Appendix B}
\noindent As mentioned in the main text, it is advantageous to work with the Laplace transform of the PGF $G(T,\xi)$, as its expression contains only a single PDF of the form given in eq.\,\ref{eq:fSPS}, rather than a convolution of $n$ PDFs. A formal proof of this result is provided here.\\
The PGF $G(T,\xi)$, defined  in the main text in terms of 
$P_T(n)$, is first rewritten as a function of the quantity $K_n(T)$, introduced in Appendix B, using the relation in eq.\,\ref{eq:PnKn}. This yields:
\begin{equation*}
G(T, \xi)= 1 +\sum_{n=1}^\infty K_n(t)\,(\xi-1)\xi^{n-1}
\end{equation*}
\noindent The Laplace transform of $G(T,\xi)$ with respect to $T$ is then evaluated. Substitution of the expression for 
$K_n(T)$ from eq.\,\ref{eq:Kn}, followed by straightforward algebraic manipulations, yields
\begin{equation*}
\tilde{G}(s,\xi)= \frac{1}{s} +\sum_{n=1}^\infty \tilde{f_n}(s)\,(\xi-1)\xi^{n-1}
\end{equation*}
\noindent At this stage, the convolution theorem is invoked, according to which the Laplace transform of a $n$-fold convolution equals the product of the $n$ Laplace transforms. Consequently, $\tilde{f_n}(s)= \left(\tilde{f} (s) \right)^n$. Substituting this result into the expression for 
$\tilde{G}(s,\xi)$ and carrying out elementary manipulations yields:
\begin{equation*}
\tilde{G}(s,\xi)= \frac{1}{s} \left\{ 1 + (\xi-1) \tilde{f}(s) \sum_{m=0}^{\infty} \left(\xi \tilde{f}(s) \right)^m \right\}
\end{equation*}
The summation in the above equation is a geometric series and admits a closed-form evaluation. Inserting this result yields the expression presented in eq.\,\ref{eq:Gtilde}.

\section*{Appendix C}
\noindent For completeness, the general expressions for the mean number of emitted photons and their variance -- derived from the PGF of a SPS under continuous-wave laser excitation and assuming ideal collection conditions -- are given by:
\begin{align*}
&\langle n \rangle_T=  \frac{\mu_1 \mu_2}{(\mu_1+\mu_2)^2}(e^{-(\mu_1+\mu_2)\,T}+(\mu_1+\mu_2)T-1)\\\\
&Var[n]_T=\frac{\mu_1 \mu_2}{(\mu_1+\mu_2 )^4}e^{-2T(\mu_1+\mu_2)}\{-\mu_1 \mu_2 + e^{2T(\mu_1+\mu_2)}[ T\mu_1^3+ \\
&\hspace{45pt}+(\mu_1^2+\mu_2^2)(-1+T\mu_2)+ \mu_1 \mu_2 (3+T\mu_2)]+ \\
&\hspace{45pt}+e^{T(\mu_1+\mu_2)}\,[\mu_2^2+\mu_1^2\,(1-4T\mu_2 )-2\mu_1 \mu_2 \,(1+ \\
&\hspace{45pt}+2T\mu_2)] \}
\end{align*}

\noindent These expressions in the long-time limit ($T\rightarrow \infty$) return eqs.\,\ref{eq:Ninfid},\ref{eq:VarNinfid}. 

\section*{Appendix D}
\noindent This section presents the transformation of eq.\,\ref{eq:fetat} into eq.\,\ref{eq:fetas}. Using the property from Appendix B that the Laplace transform of an $n$-fold convolution equals the product of the individual Laplace transforms, one has: $\tilde{f_n}(s)= \left(\tilde{f} (s) \right)^n$. Accordingly, eq.\,\ref{eq:fetat} can be rewritten as:
\begin{equation*}
\tilde{f}_\eta(s)= \sum_{n=1}^\infty (1-\eta)^{n-1}\,\eta\,(\tilde{f}(s))^n.
\end{equation*}
Through an appropriate shift of the summation index, the series can be expressed as a geometric series:
\begin{equation*}
\tilde{f}_\eta(s)= \,\eta \,f(s) \sum_{n=0}^\infty (1-\eta)^{n}\,(\tilde{f}(s))^n. 
\end{equation*}
Replacing the series with its closed-form expression results in the eq.\,\ref{eq:fetas} presented in the main text.

\section*{Authors contribution}

I.P.D., with the support of G.G. and G.P. developed the theoretical model that has been extensively and systematically analysed and discussed by I.R.B, M.G., E.M., P.T., M.Z., M.G., P.O., J.F. The manuscript was prepared with inputs by all the authors. 

\section*{Acknowledgments}

The results presented in this article had been achieved also in the context of the following projects:
QU-TEST which had received funding from the European
Union’s Horizon Europe under the grant agreement number 101113901;
PROMISE which had received funding from the European
Union’s Horizon Europe under the grant agreement number 101189611; 
DIREQT, which has received funding from the Chips Joint Undertaking (Chips JU) under grant agreement No [Project Number]; the experiment QUERIS funded by the 5th National Commission of the Italian National Institute for Nuclear Physics (INFN).
This work was also
funded by the project 23NRM04 NoQTeS, which received funding from the European Partnership on Metrology, co-financed from the European Union’s Horizon Europe Research and Innovation Programme and by the Participating States.

\section*{Data availability}
No datasets were generated or analysed during the current study.

\section*{Competing interests}
The authors declare no competing interests.

\bibliography{Bibliography}

@article{georgieva2021detection,
  title={Detection of ultra-weak laser pulses by free-running single-photon detectors: modeling dead time and dark counts effects},
  author={Georgieva, Hristina and Meda, Alice and Raupach, Sebastian MF and Hofer, Helmuth and Gramegna, Marco and Degiovanni, Ivo Pietro and Genovese, Marco and L{\'o}pez, Marco and K{\"u}ck, Stefan},
  journal={Appl. Phys. Lett.},
  volume={118},
  number={17},
  year={2021},
  doi={10.1063/5.0046014},
  publisher={AIP Publishing}
}

@article{schwartz2012improved,
  title={Improved resolution in fluorescence microscopy using quantum correlation},
  author={Schwartz, O.  and  Oron, D.},
  journal={Phys.Rev. A},
  volume={85},
  number={ },
  pages={033812},
  year={2012},
  doi={ },
  publisher={APS}
}

@article{tenne2019super,
  title={Super-resolution enhancement by quantum image scanning microscopy},
  author={Tenne, R. and Rossman, U. and  Rephael, B. and Israel, Y. and  Krupinski-Ptaszek, A. and Lapkiewicz, R. and Silberberg, Y. and Oron, D. },
  journal={Nature Photonics},
  volume={13},
  number={ },
  pages={116},
  year={2019},
  doi={ },
  publisher={Nature Publishing Group}
}

@article{Ortolano2023pattern,
  title={Quantum-Enhanced Pattern Recognition},
  author={Ortolano, Giuseppe  and Napoli, Carmine and Harney, Cillian  and Pirandola, Stefano and  Leonetti, Giuseppe  and Boucher, Pauline  and Losero, Elena and Genovese, Marco and Ruo-Berchera,  Ivano},
  journal={Phys. Rev. Applied},
  volume={20},
  number={ },
  pages={024072},
  year={2023},
  doi={ },
  publisher={APS}
}

@article{sabines2019twin,
  title={Twin-beam sub-shot-noise raster scanning microscope},
  author={Sabines-Chesterking, J.  and McMillan, A. R.  and  Moreau, P. A. and Joshi, S. K. and Knauer, S. and Johnston, E. and Rarity, J. G.  and  Matthews, J. C. F. },
  journal={Opt. Express },
  volume={27},
  number={ },
  pages={30810},
  year={2019},
  doi={ },
  publisher={Optica}
}

@article{Wang2024advantage,
  title={Quantum advantage of time-reversed ancilla based metrology of absorption parameters},
  author={Wang, J. and Filho,  R. L. d. M.  and Agarwal, G. S. and  Davidovich, L.},
  journal={Phys. Rev. Res. },
  volume={6},
  number={ },
  pages={013034 },
  year={2024},
  doi={ },
  publisher={APS}
}

@article{genovese2016real,
  title={Real applications of quantum imaging},
  author={Genovese, Marco},
  journal={Journal of Optics},
  volume={18},
  number={7},
  pages={073002},
  year={2016},
  doi={10.1088/2040-8978/18/7/073002},
  publisher={IOP Publishing}
}

@article{brida2011experimental,
  title={Experimental quantum imaging exploiting multimode spatial correlation of twin beams},
  author={Brida, Giorgio and Genovese, Marco and Ruo Berchera, Ivano},
  journal={Phys. Rev. A},
  volume={83},
  number={3},
  pages={033811},
  year={2011},
  doi={10.1103/PhysRevA.83.033811},
  publisher={APS}
}

@article{ortolano2023quantum,
  title={Quantum enhanced non-interferometric quantitative phase imaging},
  author={Ortolano, Giuseppe and Paniate, Alberto and Boucher, Pauline and Napoli, Carmine and Soman, Sarika and Pereira, Silvania F. and Ruo-Berchera, Ivano and Genovese, Marco},
  journal={Light: Science \& Applications},
  volume={12},
  pages={171},
  year={2023},
  doi={10.1038/s41377-023-01215-1},
  publisher={Nature Publishing Group}
}

@article{gattomonticone2014native,
  title={Native NIR-emitting single colour centres in CVD diamond},
  author={Gatto Monticone, D. and Traina, P. and Moreva, E. and Forneris, J. and Olivero, P. and Degiovanni, I. P. and Taccetti, F. and Giuntini, L. and Brida, G. and Amato, G. and Genovese, M.},
  journal={New Journal of Physics},
  volume={16},
  number={5},
  pages={053005},
  year={2014},
  doi={10.1088/1367-2630/16/5/053005},
  publisher={IOP Publishing}
}

@article{gattomonticone2014beating,
  title={Beating the Abbe Diffraction Limit in Confocal Microscopy via Nonclassical Photon Statistics},
  author={Gatto Monticone, D. and Katamadze, K. and Traina, P. and Moreva, E. and Forneris, J. and Ruo-Berchera, I. and Olivero, P. and Degiovanni, I. P. and Brida, G. and Genovese, M.},
  journal={Phys. Rev. Lett.},
  volume={113},
  number={14},
  pages={143602},
  year={2014},
  doi={10.1103/PhysRevLett.113.143602},
  publisher={APS}
}

@article{sipahigil2014indistinguishable,
  title={Indistinguishable photons from separated silicon-vacancy centers in diamond},
  author={Sipahigil, Alp and Jahnke, Kay D and Rogers, Lachlan J and Teraji, Tokuyuki and Isoya, Junichi and Zibrov, Alexander S and Jelezko, Fedor and Lukin, Mikhail D},
  journal={Phys. Rev. Lett.},
  volume={113},
  number={11},
  pages={113602},
  year={2014},
  doi={10.1103/PhysRevLett.113.113602},
  publisher={APS}
}

@article{storteboom2015lifetime,
  title={Lifetime investigation of single nitrogen vacancy centres in nanodiamonds},
  author={Storteboom, Jelle and Dolan, Philip and Castelletto, Stefania and Li, Xiangping and Gu, Min},
  journal={Opt. Express},
  volume={23},
  number={9},
  pages={11327--11333},
  year={2015},
  doi={10.1364/OE.23.011327},
  publisher={Optical Society of America}
}

@article{nguyen2019photodynamics,
  title={Photodynamics and quantum efficiency of germanium vacancy color centers in diamond},
  author={Nguyen, Minh and Nikolay, Niko and Bradac, Carlo and Kianinia, Mehran and Ekimov, Evgeny A and Mendelson, Noah and Benson, Oliver and Aharonovich, Igor},
  journal={Adv. Photon.},
  volume={1},
  number={6},
  pages={066002--066002},
  year={2019},
  doi={10.1117/1.AP.1.6.066002},
  publisher={Society of Photo-Optical Instrumentation Engineers}
}

@article{berchera2019quantum,
  title={Quantum imaging with sub-Poissonian light: challenges and perspectives in optical metrology},
  author={Berchera, I Ruo and Degiovanni, Ivo Pietro},
  journal={Metrologia},
  volume={56},
  number={2},
  pages={024001},
  year={2019},
  doi={10.1088/1681-7575/aaf7b2},
  publisher={IOP Publishing}
}

@article{picariello2025quantum,
  title={Quantum super-resolution microscopy by photon statistics and structured light},
  author={Picariello, Fabio and Losero, Elena and Ditalia Tchernij, Sviatoslav and Boucher, Pauline and Genovese, Marco and Ruo-Berchera, Ivano and Degiovanni, Ivo Pietro},
  journal={Optica},
  volume={12},
  number={4},
  pages={490--497},
  year={2025},
  doi={10.1364/OPTICA.540264},
  publisher={Optica Publishing Group}
}

@article{costa2026quantum,
  title={Quantum dots on GaAs substrates as integration-ready high-performance single-photon sources at telecommunication wavelengths},
  author={Costa, Beatrice and Scaparra, Bianca and Wei, Xiao and Riedl, Hubert and Koblm{\"u}ller, Gregor and Zallo, Eugenio and Finley, Jonathan J and Hanschke, Lukas and M{\"u}ller, Kai},
  journal={Phys. Rev. Appl.},
  volume={25},
  number={1},
  pages={L011002},
  year={2026},
  doi={10.1103/4fzb-595x},
  publisher={APS}
}

@article{corte2022magnesium,
  title={Magnesium-vacancy optical centers in diamond},
  author={Corte, Emilio and Andrini, Greta and Nieto Hern{\'a}ndez, Elena and Pugliese, Vanna and Costa, {\^A}ngelo and Magchiels, Goele and Moens, Janni and Tunhuma, Shandirai Malven and Villarreal, Renan and Pereira, Lino MC and others},
  journal={ACS Photonics},
  volume={10},
  number={1},
  pages={101--110},
  year={2022},
  doi={10.1021/acsphotonics.2c01130},
  publisher={ACS Publications}
}

@article{parto2021defect,
  title={Defect and strain engineering of monolayer WSe2 enables site-controlled single-photon emission up to 150 K},
  author={Parto, Kamyar and Azzam, Shaimaa I and Banerjee, Kaustav and Moody, Galan},
  journal={Nat. commun.},
  volume={12},
  number={1},
  pages={3585},
  year={2021},
  doi={10.1038/s41467-021-23709-5},
  publisher={Nature Publishing Group UK London}
}

@article{palm2023modular,
  title={Modular chip-integrated photonic control of artificial atoms in diamond waveguides},
  author={Palm, Kevin J and Dong, Mark and Golter, D Andrew and Clark, Genevieve and Zimmermann, Matthew and Chen, Kevin C and Li, Linsen and Menssen, Adrian and Leenheer, Andrew J and Dominguez, Daniel and others},
  journal={Optica},
  volume={10},
  number={5},
  pages={634--641},
  year={2023},
  doi={10.1364/OPTICA.486361},
  publisher={Optica Publishing Group}
}

@article{grosso2017tunable,
  title={Tunable and high-purity room temperature single-photon emission from atomic defects in hexagonal boron nitride},
  author={Grosso, Gabriele and Moon, Hyowon and Lienhard, Benjamin and Ali, Sajid and Efetov, Dmitri K and Furchi, Marco M and Jarillo-Herrero, Pablo and Ford, Michael J and Aharonovich, Igor and Englund, Dirk},
  journal={Nat. commun.},
  volume={8},
  number={1},
  pages={1--8},
  year={2017},
  doi={10.1038/s41467-017-00810-2},
  publisher={Nature Publishing Group}
}

@article{brambilla2008high,
  title={High-sensitivity imaging with multi-mode twin beams},
  author={Brambilla, Elena and Caspani, L and Jedrkiewicz, Ottavia and Lugiato, LA and Gatti, A},
  journal={Phys. Rev. A},
  volume={77},
  number={5},
  pages={053807},
  year={2008},
  doi={10.1103/PhysRevA.77.053807},
  publisher={APS}
}

@article{brida2010experimental,
  title={Experimental realization of sub-shot-noise quantum imaging},
  author={Brida, Giorgio and Genovese, Marco and Ruo Berchera, Ivano},
  journal={Nat. Photon.},
  volume={4},
  number={4},
  pages={227--230},
  year={2010},
  doi={10.1038/nphoton.2010.29},
  publisher={Nature Publishing Group UK London}
}

@article{radulaski2017scalable,
  title={Scalable quantum photonics with single color centers in silicon carbide},
  author={Radulaski, Marina and Widmann, Matthias and Niethammer, Matthias and Zhang, Jingyuan Linda and Lee, Sang-Yun and Rendler, Torsten and Lagoudakis, Konstantinos G and Son, Nguyen Tien and Janzen, Erik and Ohshima, Takeshi and others},
  journal={Nano lett.},
  volume={17},
  number={3},
  pages={1782--1786},
  year={2017},
  doi={10.1021/acs.nanolett.6b05102},
  publisher={ACS Publications}
}

@article{lounis2005single,
  title={Single-photon sources},
  author={Lounis, Brahim and Orrit, Michel},
  journal={Rep. Prog. Phys.},
  volume={68},
  number={5},
  doi={10.1088/0034-4885/68/5/R04},
  pages={1129--1179},
  year={2005}
}

@article{higginbottom2016pure,
  title={Pure single photons from a trapped atom source},
  author={Higginbottom, Daniel B and Slodi{\v{c}}ka, Luk{\'a}{\v{s}} and Araneda, Gabriel and Lachman, Luk{\'a}{\v{s}} and Filip, Radim and Hennrich, Markus and Blatt, Rainer},
  journal={New J. Phys.},
  volume={18},
  number={9},
  pages={093038},
  year={2016},
  doi={10.1088/1367-2630/18/9/093038},
  publisher={IOP Publishing}
}

@article{samantaray2017realization,
  title={Realization of the first sub-shot-noise wide field microscope},
  author={Samantaray, Nigam and Ruo-Berchera, Ivano and Meda, Alice and Genovese, Marco},
  journal={Light: Sci. Appl.},
  volume={6},
  number={7},
  pages={e17005--e17005},
  year={2017},
  doi={10.1038/lsa.2017.5},
  publisher={Nature Publishing Group}
}

@article{ruo2020improving,
  title={Improving resolution-sensitivity trade off in sub-shot noise quantum imaging},
  author={Ruo-Berchera, Ivano and Meda, Alice and Losero, Elena  and Avella, Alessio and Samantaray, Nigam and Genovese, Marco},
  journal={Applied Physics Letters},
  volume={116},
  number={21},
  pages={214001},
  year={2020},
  doi={ },
  publisher={APS }
}

@article{paniate2026high,
  title={High-resolution quantum-enhanced phase imaging of cells},
  author={Paniate, Alberto and Ortolano Giuseppe and Soman, S. and Genovese, Marco and Ruo-Berchera, Ivano },
  journal={Optica},
  volume={13},
  number={  },
  pages={375-385},
  year={2026},
  doi={ },
  publisher={Optica }
}

@article{Lawrie2019QuantumSensing,
  author  = {Lawrie, B. J. and Lett, P. D. and Marino, A. M. and Pooser, R. C.},
  title   = {Quantum Sensing with Squeezed Light},
  journal = {ACS Photonics},
  year    = {2019},
  volume  = {6},
  number  = {6},
  pages   = {1307--1318},
  doi     = {10.1021/acsphotonics.9b00250},
  url     = {https://doi.org/10.1021/acsphotonics.9b00250}
}

@article{Defienne2024AdvancesQuantumImaging,
  author  = {Defienne, Hugo and Bowen, Warwick P. and Chekhova, Maria and
             Barreto Lemos, Gabriela and Oron, Dan and Ramelow, Sven and
             Treps, Nicolas and Faccio, Daniele},
  title   = {Advances in Quantum Imaging},
  journal = {Nature Photonics},
  year    = {2024},
  volume  = {18},
  number  = {10},
  pages   = {1024--1036},
  doi     = {10.1038/s41566-024-01516-w},
  url     = {https://doi.org/10.1038/s41566-024-01516-w}
}

@article{Pirandola2018AdvancesPhotonicSensing,
  author  = {Pirandola, Stefano and Bardhan, Bhaskar Roy and Gehring, Tobias and
             Weedbrook, Christian and Lloyd, Seth},
  title   = {Advances in Photonic Quantum Sensing},
  journal = {Nature Photonics},
  year    = {2018},
  volume  = {12},
  number  = {12},
  pages   = {724--733},
  doi     = {10.1038/s41566-018-0301-6},
  url     = {https://doi.org/10.1038/s41566-018-0301-6}
}

@article{Dorfman2016NonlinearOpticalSignals,
  author  = {Dorfman, Konstantin E. and Schlawin, Frank and Mukamel, Shaul},
  title   = {Nonlinear Optical Signals and Spectroscopy with Quantum Light},
  journal = {Reviews of Modern Physics},
  year    = {2016},
  volume  = {88},
  number  = {4},
  pages   = {045008},
  doi     = {10.1103/RevModPhys.88.045008},
  url     = {https://doi.org/10.1103/RevModPhys.88.045008}
}

@article{Berthel2015Photophysics,
  author  = {Berthel, M. and Mollet, Oriane and Dantelle, G{\'e}raldine and
             Gacoin, Thierry and Huant, Serge and Drezet, Aur{\'e}lien},
  title   = {Photophysics of Single Nitrogen-Vacancy Centers in Diamond Nanocrystals},
  journal = {Physical Review B},
  year    = {2015},
  volume  = {91},
  number  = {3},
  pages   = {035308},
  doi     = {10.1103/PhysRevB.91.035308}
}

@article{Santori2004SinglePhoton,
  author  = {Santori, Charles and Fattal, David and Vu{\v{c}}kovi{\'c}, Jelena and
             Solomon, Glenn S. and Yamamoto, Yoshihisa},
  title   = {Single-Photon Generation with {InAs} Quantum Dots},
  journal = {New Journal of Physics},
  year    = {2004},
  volume  = {6},
  number  = {1},
  pages   = {89},
  doi     = {10.1088/1367-2630/6/1/089}
}

@article{Demissie2020TripletShelving,
  author  = {Demissie, Aida A. and Dickson, Robert M.},
  title   = {Triplet Shelving in Fluorescein and Its Derivatives Provides
             Delayed, Background-Free Fluorescence Detection},
  journal = {The Journal of Physical Chemistry A},
  year    = {2020},
  volume  = {124},
  number  = {7},
  pages   = {1437--1443},
  doi     = {10.1021/acs.jpca.9b11040}
}

@book{Fox2010OpticalProperties,
  author    = {Fox, Mark},
  title     = {Optical Properties of Solids},
  publisher = {Oxford University Press},
  address   = {Oxford},
  year      = {2010},
  edition   = {2},
  series    = {Oxford Master Series in Physics},
  volume    = {3},
  isbn      = {978-0-19-957336-3}
}

@article{Alkauskas2014Luminescence,
  author  = {Alkauskas, Audrius and Buckley, Benjamin B. and
             Awschalom, David D. and Van de Walle, Chris G.},
  title   = {First-Principles Theory of the Luminescence Lineshape for the
             Triplet Transition in Diamond {NV} Centres},
  journal = {New Journal of Physics},
  year    = {2014},
  volume  = {16},
  number  = {7},
  pages   = {073026},
  doi     = {10.1088/1367-2630/16/7/073026}
}

@article{Mandel1979SubPoissonian,
  author  = {Mandel, Leonard},
  title   = {Sub-Poissonian Photon Statistics in Resonance Fluorescence},
  journal = {Optics Letters},
  year    = {1979},
  volume  = {4},
  number  = {6},
  pages   = {205--207},
  doi     = {10.1364/OL.4.000205}
}

@article{Hodapp1990,
  author  = {Hodapp, T. W. and Greenlees, G. W. and Finn, M. A. and Lewis, D. A.},
  title   = {Resonant light scattering from single atoms: Sub-Poissonian statistics versus antibunching},
  journal = {Physical Review A},
  volume  = {41},
  number  = {5},
  pages   = {2698},
  year    = {1990},
  doi     = {10.1103/PhysRevA.41.2698}
}

@article{Fano1947,
  author  = {Fano, U.},
  title   = {Ionization Yield of Radiations. II. The Fluctuations of the Number of Ions},
  journal = {Physical Review},
  volume  = {72},
  number  = {1},
  pages   = {26--29},
  year    = {1947},
  doi     = {10.1103/PhysRev.72.26}
}

@book{cox1962renewal,
  author    = {Cox, David R.},
  title     = {Renewal Theory},
  series    = {Monographs on Applied Probability and Statistics},
  publisher = {Methuen \& Co.},
  address   = {London},
  year      = {1962},
  isbn      = {978-0416523805}
}

@article{EsmannWeinAntonSolanas2024SolidState,
  author  = {Esmann, Martin and Wein, Stephen C. and Ant{\'o}n-Solanas, Carlos},
  title   = {Solid-State Single-Photon Sources: Recent Advances for Novel Quantum Materials},
  journal = {Advanced Functional Materials},
  year    = {2024},
  volume  = {34},
  number  = {30},
  pages   = {2315936},
  doi     = {10.1002/adfm.202315936}
}

@article{Mohammadnejad2023RoomTemperature,
  author  = {Mohammadnejad, Shahram and Nosratkhah, Pouya and Arab, Hossein},
  title   = {Recent advances in room temperature single-photon emitters},
  journal = {Quantum Information Processing},
  year    = {2023},
  volume  = {22},
  number  = {10},
  pages   = {360},
  doi     = {10.1007/s11128-023-04100-3}
}

@article{Holmes2020QuantumDotSources,
  author  = {Holmes, M. J. H. and Arita, M. and Arakawa, Y.},
  title   = {Progress in quantum-dot single photon sources for quantum information technologies: A broad spectrum overview},
  journal = {Applied Physics Reviews},
  year    = {2020},
  volume  = {7},
  number  = {2},
  pages   = {021309},
  doi     = {10.1063/5.0010193}
}

@article{Senellart2017HighPerformanceQD,
  author  = {Senellart, Pascale and Solomon, Glenn and White, Andrew},
  title   = {High-performance semiconductor quantum-dot single-photon sources},
  journal = {Nature Nanotechnology},
  year    = {2017},
  volume  = {12},
  number  = {11},
  pages   = {1026--1039},
  doi     = {10.1038/nnano.2017.218}
}

@article{Andrini2024SolidStateColorCenters,
  author  = {Andrini, Greta and others},
  title   = {Solid-State Color Centers for Single-Photon Generation},
  journal = {Photonics},
  year    = {2024},
  volume  = {11},
  number  = {2},
  pages   = {188},
  doi     = {10.3390/photonics11020188}
}

@article{Castelletto2021SiCSources,
  author  = {Castelletto, Stefania},
  title   = {Silicon carbide single-photon sources: challenges and prospects},
  journal = {Materials for Quantum Technology},
  year    = {2021},
  volume  = {1},
  number  = {2},
  pages   = {023001},
  doi     = {10.1088/2633-4356/abe04a}
}

@article{TothAharonovich2019AtomicallyThin,
  author  = {Toth, Milos and Aharonovich, Igor},
  title   = {Single Photon Sources in Atomically Thin Materials},
  journal = {Annual Review of Physical Chemistry},
  year    = {2019},
  volume  = {70},
  pages   = {123--142},
  doi     = {10.1146/annurev-physchem-042018-052628}
}

@article{Zhang2024TwoDimensionalEmitters,
  author  = {Zhang, Chengzhi and Gong, Zehuizi and He, Dawei and Yan, Yige and Li, Songze and Zhao, Kun and Wang, Jiarong and Wang, Yongsheng and Zhang, Xiaoxian},
  title   = {Research Progress of Single-Photon Emitters Based on Two-Dimensional Materials},
  journal = {Nanomaterials},
  year    = {2024},
  volume  = {14},
  number  = {11},
  pages   = {918},
  doi     = {10.3390/nano14110918}
}

@article{Ye2024ColloidalQDs,
  author  = {Ye, Yongzheng and Lin, Xing and Fang, Wei},
  title   = {Room-Temperature Single-Photon Sources Based on Colloidal Quantum Dots: A Review},
  journal = {Materials},
  year    = {2023},
  volume  = {16},
  number  = {24},
  pages   = {7684},
  doi     = {10.3390/ma16247684}
}

@inproceedings{Rempe2007CavityQED,
  author    = {Rempe, G. and Hijlkema, M. and Kuhn, A. and Specht, H. P. and Weber, B. and Webster, S. and Wilk, T.},
  title     = {Optical Cavity Quantum Electrodynamics: A Toolbox for Deterministic Quantum Information Science with Stationary Atoms and Flying Photons},
  booktitle = {Conference on Coherence and Quantum Optics},
  year      = {2007},
  pages     = {117--127},
  publisher = {Optica Publishing Group},
  note      = {OSA Technical Digest, paper CMF3}
}

@article{DuanMonroe2010TrappedIonNetworks,
  author  = {Duan, L.-M. and Monroe, Christopher},
  title   = {Colloquium: Quantum networks with trapped ions},
  journal = {Reviews of Modern Physics},
  year    = {2010},
  volume  = {82},
  number  = {2},
  pages   = {1209--1224},
  doi     = {10.1103/RevModPhys.82.1209}
}

@article{Toninelli2021SingleMolecules,
  author  = {Toninelli, C. and Gerhardt, I. and Clark, A. S. and Reserbat-Plantey, A. and G{\"o}tzinger, S. and Ristanovi{\'c}, Z. and Colautti, M. and Lombardi, P. and Major, K. D. and Deperasi{\'n}ska, I. and Pernice, W. H. and Koppens, F. H. L. and Kozankiewicz, B. and Gourdon, A. and Sandoghdar, V. and Orrit, M.},
  title   = {Single organic molecules for photonic quantum technologies},
  journal = {Nature Materials},
  year    = {2021},
  volume  = {20},
  number  = {12},
  pages   = {1615--1628},
  doi     = {10.1038/s41563-021-00987-4}
}

@article{ShkarinGotzinger2026Molecules,
  author  = {Shkarin, Alexey and G{\"o}tzinger, Stephan},
  title   = {Organic molecules as single-photon sources},
  journal = {Applied Physics Reviews},
  year    = {2026},
  volume  = {13},
  number  = {2},
  pages   = {021312},
  doi     = {10.1063/5.0304253}
}
\end{document}